\documentclass[aps, prb, reprint, twocolumn, superscriptaddress, letterpaper, longbibliography, 10pt, floatfix, showpacs, balancelastpage]{revtex4-2}
\usepackage{graphicx, float, ulem, mathrsfs, braket, bbold, bm, color, psfrag, amsmath, amssymb, dcolumn, mathrsfs}
\pdfoutput=1
\usepackage[mathlines]{lineno}
\usepackage[colorlinks = True, citecolor = blue, linkcolor = blue, allcolors = blue]{hyperref}
\usepackage{enumitem, array, makecell}
\usepackage[utf8]{inputenc}
\usepackage{orcidlink, tikz}
\usepackage[linesnumbered,ruled,vlined]{algorithm2e}
\usepackage{algpseudocode}

\def\bea{\begin{eqnarray}}
\def\eea{\end{eqnarray}}
\def\bal{\begin{aligned}}
\def\eal{\end{aligned}}

\begin{document}
\title{Probing the limits of variational quantum algorithms for nonlinear ground states on real quantum hardware: The effects of noise}
\author{Muhammad Umer\orcidlink{0000-0002-1941-1833}}
\email{umer@u.nus.edu}
\affiliation{Centre for Quantum Technologies, National University of Singapore, 3 Science Drive 2, Singapore 117543}
\author{Eleftherios Mastorakis\orcidlink{0009-0005-3546-2568}}
\affiliation{School of Electrical and Computer Engineering, Technical University of Crete, Chania, Greece 73100}
\author{Sofia Evangelou}
\affiliation{School of Electrical and Computer Engineering, Technical University of Crete, Chania, Greece 73100}
\author{Dimitris G. Angelakis\orcidlink{0000-0001-6763-6060}}
\email{daggelakis@tuc.gr}
\affiliation{Centre for Quantum Technologies, National University of Singapore, 3 Science Drive 2, Singapore 117543}
\affiliation{School of Electrical and Computer Engineering, Technical University of Crete, Chania, Greece 73100}
\affiliation{AngelQ Quantum Computing, 531A Upper Cross Street, \#04-95 Hong Lim Complex, Singapore 051531}
\date{\today}

\begin{abstract}
A recently proposed variational quantum algorithm has expanded the horizon of variational quantum computing to nonlinear physics and fluid dynamics. In this work, we probe the ability of such approaches to capture the ground state of the nonlinear Schr\"{o}dinger equation for a range of parameters on real superconducting quantum processors. Specifically, we study the expressivity of real-amplitude, hardware-efficient ansatz to capture the ground state of this nonlinear system across various interaction regimes and implement different noise scenarios in both simulators and cloud processors. Our investigation reveals that although quantum hardware noise impairs the evaluation of the energy cost function, certain small instances of the problem consistently converge to the ground state. We test for a variety of cases on IBM Q superconducting devices and analyze the discrepancies in the energy cost function evaluation due to quantum hardware noise. These discrepancies are absent in the state fidelity estimation because of the shallow state preparation circuit. Our comprehensive analysis offers valuable insights into the practical implementation and advancement of the variational algorithms for nonlinear problems.
\end{abstract}
\maketitle

\section{Introduction}\label{Sec:Introduction}
\par Quantum computing has gained significant attention in past decades with the potential to efficiently solve classically intractable problems. In this regard, various quantum algorithms have been proposed to use quantum devices for computation. Such algorithms include but are not limited to, Shor's algorithm \cite{Shor1994} for efficiently factorizing a composite number into its prime factors, Grover's algorithm \cite{Grover1996} for searching unsorted databases, Harrow-Hassidim-Lloyd (HHL) algorithm \cite{Harrow2009} for solving systems of linear equations, and quantum simulation algorithms \cite{Abrams1997, *Abrams1999} for simulating quantum systems. These quantum algorithms promise performance, in terms of solving the above-mentioned problems efficiently, unparalleled by any classical algorithm to date for the same problem. Unfortunately, the quantum hardware requirements of these algorithms are far beyond the current capabilities of the noisy intermediate-scale quantum (NISQ) devices \cite{Cerezo2021, Bharti2022}. 

\par In light of the constrained capabilities of NISQ devices, variational quantum algorithms (VQAs) \cite{Cerezo2021, Peruzzo2014, Kandala2017, McClean2016, Bharti2022} have garnered considerable traction over the past decade. VQAs are hybrid quantum-classical algorithms designed to utilize a quantum device to approximately construct the solution to a problem through a parameterized quantum circuit (PQC), the parameters of which are optimized using a classical computer. The parameterized quantum circuit (PQC) generates a quantum state, and the expectation value of an observable related to the problem is evaluated, serving as the cost function for the classical optimizer, whose minimum/maximum value represents the solution to the problem at hand. VQAs exhibit a broad range of applications that are ever-expanding and span areas such as quantum chemistry \cite{Peruzzo2014, Kandala2017, OMalley2016, Hempel2018, Ganzhorn2019, Quantum2020} for material and drug discovery, addressing combinatorial optimization problems \cite{Farhi2014, Pagano2020, Tan2021, Zhu2022} in logistics and finance, and handling various machine learning tasks \cite{Benedetti2019, Zhu2019, Tangpanitanon2020}. 

\par
\begin{figure}[hbt]\begin{center}
\includegraphics[clip, trim=0.00cm 0.00cm 0.05cm 0.05cm, width=0.95\linewidth, height=0.55\linewidth, angle=0]{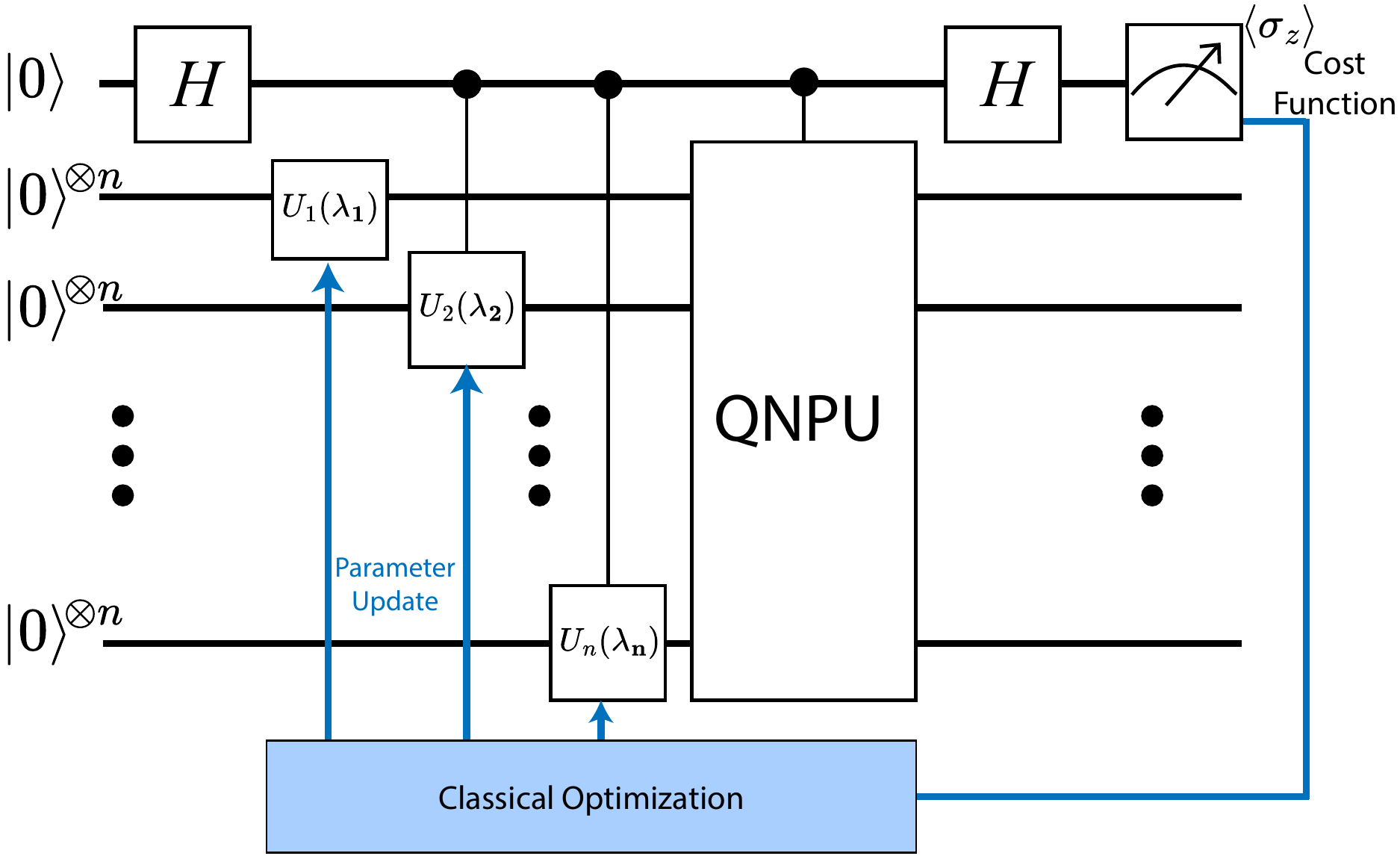}
\caption{Schematic depiction of a variational quantum algorithm for nonlinear problems, adapted from Ref. \cite{Lubasch2020}. Here, $H$ is the Hadamard operator, $\vert{0}\rangle$ ($\vert{0}\rangle^{\otimes{n}}$) is the single- (multi-) qubit initial state, $U_{i}(\lambda_{i})$ is the unitary operator representing an ansatz that prepares a trial state on the ${i}^{\rm th}$ quantum register, and QNPU is the quantum nonlinear processing unit that evaluates the nonlinear cost function. The value of the cost function is measured by evaluating the Pauli $\sigma_{z}$ expectation value on the ancilla qubit. This cost function value is then fed into the classical optimizer, which updates the variational parameters of the ansatz until convergence is achieved. \vspace{-0.5cm}}
\label{Fig:Algorithm}
\end{center}\end{figure}

 Recently, the domain of VQAs has been further expanded to encompass applications in computational fluid dynamics (CFD) \cite{Jaksch2023} and other nonlinear physics areas \cite{Lubasch2020, Sarma2024}, which paves the way for solving complex nonlinear problems using quantum devices. A generic schematic, shown in Fig. \ref{Fig:Algorithm}, highlights the structure and operational flow of the variational algorithm for nonlinear problems. In addition to the conventional building blocks of VQAs, such as initial state encoding, parameterized quantum ansatz, and classical optimization, VQAs for nonlinear problems have an additional component: the quantum nonlinear processing unit (QNPU), which accepts multiple trial states (which may be identical or otherwise) to evaluate a nonlinear cost function, as depicted in Fig. \ref{Fig:Algorithm}. It is important to note that while Refs. \cite{Lubasch2020, Jaksch2023} introduce the variational quantum algorithm to study ground state problems and time evolution of nonlinear systems on quantum devices, these works do not demonstrate the practical implementation of such algorithms, either on simulators or quantum hardware. In this study, we address this gap by conducting a comprehensive analysis of the variational approach to nonlinear problems, both in the presence and absence of hardware noise, and by implementing the algorithm on quantum devices.

\par In this article, our objective is twofold. First, we aim to analyze the performance of the variational quantum algorithm in solving the ground state problem of a one-dimensional time-independent nonlinear Schr\"{o}dinger equation (NLSE). We report that real-amplitude ansatz captures the ground state of the problem across various regimes characterized by varying strengths of nonlinearity. Our second objective is to investigate the effects of hardware noise on the performance of the variational quantum algorithm. For this purpose, we first incorporate the hardware noise in simulations to solve for the ground state of the NLSE within the strong nonlinearity regime, for a small system size. We then implement pre-trained quantum circuits on digital, gate-based superconducting devices. Our findings reveal that hardware noise affects the execution of various blocks within the variational quantum algorithm, leading to discrepancies in the cost function values. We observe a substantial overlap between the trial states constructed in noisy and noiseless settings, demonstrating that the disparities in the cost function values arise from the computation processes executed on the variational state. Furthermore, we reveal that despite the discrepancies in the cost function values, the algorithm converges to a set of parameters that captures the ground state with over $99\%$ fidelity. Our investigation underscores the adaptability of the variational algorithm in solving complex nonlinear problems, concurrently highlighting the limitations imposed by the hardware noise inherent in the NISQ devices. 

\par The rest of the article is structured as follows: Sec. \ref{Sec:NLSE_example} introduces the ground state problem of the time-independent nonlinear Schr\"{o}dinger equation (NLSE) and discusses its implementation using the variational quantum approach. Sec. \ref{Sec:Results} presents numerical results for solving the ground state problem of the NLSE, first in the absence of quantum noise and then in its presence, detailed in Sec. \ref{Sec:Noiseless_Simulations} and Sec. \ref{Sec:Noisy_Simulations}, respectively. Furthermore, Sec. \ref{Sec:Device_Demonstration} discusses the quantum hardware implementation of variational quantum circuits. We summarize in Sec. \ref{Sec_Conclusion} and also present an outlook for future research directions.

\section{Nonlinear Schr\"{o}dinger equation on a Quantum Computer} 
\label{Sec:NLSE_example}
 
\par The nonlinear Schr\"{o}dinger equation (NLSE) and its variants model various phenomena \cite{Scott2006, Agrawal2013, Nakkeeran2002, Triki2019, Sulem1999, Gross1961, Pitaevskii1961, Dalfovo1999, Leggett2001, Pitaevskii2003}, such as dynamics of light in nonlinear optics \cite{Nakkeeran2002, Triki2019}, envelope solitons and modulation instabilities in plasma physics and surface gravity waves \cite{Sulem1999}, and characteristics including superfluidity and vortex formation in Bose-Einstein Condensates (BEC) \cite{Gross1961, Pitaevskii1961, Dalfovo1999, Leggett2001, Pitaevskii2003}, to name a few. In dimensionless form, the time-independent NLSE is given as 
\bea\bal
\big[-\frac{1}{2}\frac{d^{2}}{d{x}^{2}} + V(x) + g{I_{f(x)}} \big] {f(x)} = E {f(x)}\;.
\label{EQ:Schrodinger}
\eal\eea 
Here, $f(x)$, with $x$ being spatial coordinates, represents a normalized single real-valued function defined over the interval $[a, b]$. The term $I_{f(x)}$ represents the nonlinear interaction, $g$ denotes the strength of the nonlinearity, and $V(x)$ is the external potential. In this study, we consider $I_{f(x)} = \vert{f(x)}\vert^{2}$, quadratic potential $V(x) = V_{0}(x - x_{0})^{2}$ centered around $x_{0} = \frac{b-a}{2}$, and periodic boundary conditions such that $f(b) = f(a)$ and $V(b) = V(a)$. Small instances of the Eq. (\ref{EQ:Schrodinger}) can be solved numerically on classical computers by employing imaginary-time evolution \cite{Edwards1995, Dalfovo1999, Lubasch2018}, spectral, variational or other methods \cite{Scott2006}. However, when addressing large instances of nonlinear problems, the limitations associated with memory capacity and computational time inherent in classical computation become increasingly apparent. It is worth highlighting that Ref. \cite{Lubasch2020} discusses the procedure and quantum circuits for implementing the ground state problem of the NLSE on quantum computers, but it does not analyze the performance of the algorithm in either simulations or on quantum hardware. We revisit this procedure below for completeness, followed by a thorough numerical analysis in the subsequent sections.

\begin{figure}[hbt]\begin{center}
\includegraphics[clip, trim=0.0cm 0.68cm 0.0cm 0.0cm, width=0.98\linewidth, height=1.60\linewidth, angle=0]{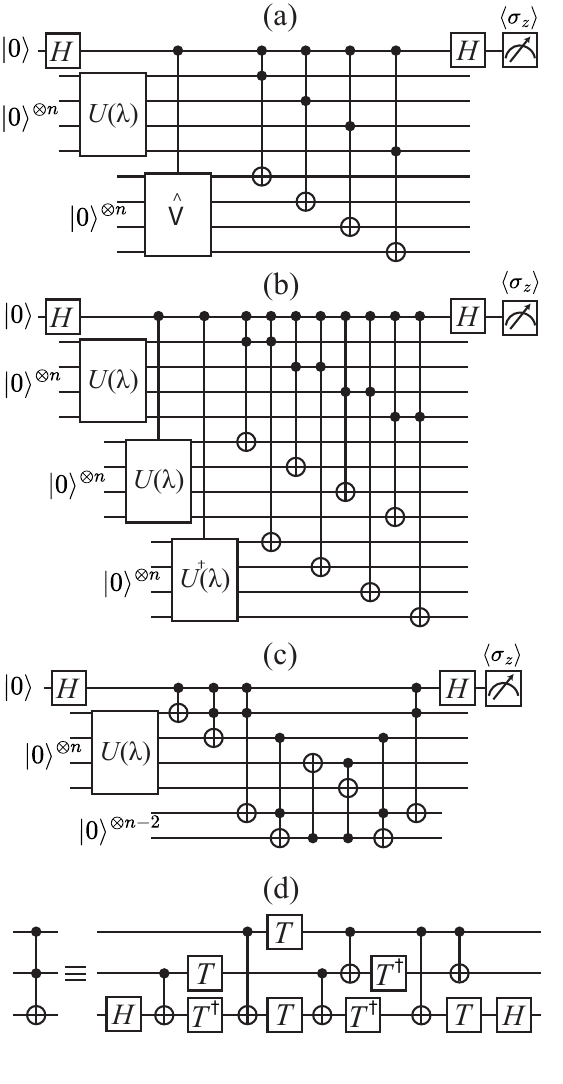}
\caption{Design of the quantum circuits used to measure the (a) potential, (b) interaction, and (c) kinetic energies. Here, $H$ is the Hadamard gate, $U(\lambda)$ is the quantum ansatz that represents the trial state, and $\hat{V}$ is the potential unitary which encodes the potential function values to the basis states. Here, we have shown an example of $n = 4$ ($n' = 13$), which can be generalized to an arbitrary number of qubits. Panel (d) shows the decomposition of the Toffoli gate into a sequence of single-qubit gates and controlled-NOT gates, where $ T = R_{z}(\pi/4)$ and $T^{\dagger} = R_{z}(-\pi/4)$. \vspace{-0.2cm}}
\label{Fig:QNPU}
\end{center}\end{figure}

\par Following standard numerical approaches, we discretize the interval $[a, b]$ into $N$ equidistant points $x_{k} = a + \delta{k}$, where $\delta = (b-a)/N$ is the spacing between two adjacent grid points, and $k \in \{0, 1, 2, \dots, N-1\}$. The normalization condition on the function $f(x)$ takes the form $1 = \delta\sum_{k = 0}^{N-1} \vert{f(x_k)}\vert^{2} = \sum_{k = 0}^{N-1} \vert{\psi_{k}}\vert^{2}$, where we have defined $\psi_{k} = \sqrt{\delta}f(x_{k})$. We encode the $N = 2^{n}$ amplitudes $\psi_{k}$, which may also incorporate the initial conditions of the problem, into the basis states $\vert{\rm binary}(k)\rangle$ of the $n$-qubit quantum register such that the quantum state takes the form $\vert\Psi\rangle = \sum_{k}\psi_{k}\vert{\rm binary}(k)\rangle$. By applying the finite-difference method, the expectation value of the total energy [from Eq. (\ref{EQ:Schrodinger})] of the system is given as the sum of potential, interaction, and kinetic energies, $\langle\langle{E}\rangle\rangle = \langle\langle{E_{P}}\rangle\rangle + \langle\langle{E_{I}}\rangle\rangle + \langle\langle{E_{K}}\rangle\rangle$, where
\bea\bal
\label{EQ:Cost_Functions}
\langle\langle{E_{P}}\rangle\rangle &= \sum_{k = 0}^{N-1}~\vert\psi_{k}\vert^{2}V_{k}\;,~~~~ \langle\langle{E_{I}}\rangle\rangle = \sum_{k = 0}^{N-1}~\frac{g}{\delta}\vert\psi_{k}\vert^{4}\;,\\
\langle\langle{E_{K}}\rangle\rangle &= -\frac{1}{2\delta^{2}}\sum_{k = 0}^{N-1}\big(\psi^{\ast}_{k}\psi_{k+1} - 2\psi^{\ast}_{k}\psi_{k} + \psi^{\ast}_{k}\psi_{k-1}\big)\;,
\eal\eea
for the discretized problem and $\langle\langle{\cdots}\rangle\rangle$ represent the expectation value \cite{Lubasch2020}. We consider the total energy as the cost function $\mathcal{C} = \sum_{j}\mathcal{C}_{j} = \sum_{j}\langle\langle{E_{j}}\rangle\rangle$ for the variational algorithm such that the minimum value of the cost function represents the ground state solution. 

\par Each component of the energy cost function is evaluated separately and requires a dedicated quantum nonlinear processing unit (QNPU), as shown in Figs. \ref{Fig:QNPU}(a-c), which are adapted from Ref. \cite{Lubasch2020}. For the measurement of potential (interaction) energy, the relevant QNPU constructs the potential function (variational states) on the separate quantum register(s) and performs its (their) bit-wise multiplication with the primary variational state, as depicted in Fig. \ref{Fig:QNPU}(a) (Fig. \ref{Fig:QNPU}(b)). The procedure of encoding the potential values in the unitary operator $\hat{V}$ is well known in the quantum computing literature \cite{Oseledets2013} and is elaborated in Appendix \ref{App_Sec:MPS}. The kinetic energy is calculated using an adder circuit \cite{Vedral1996, Nielsen2010, Lubasch2020}, which requires an additional $n-2$ ancilla qubits, illustrated in Fig. \ref{Fig:QNPU}(c), where $n$ represents the number of qubits in the primary quantum register. It is important to highlight that performing the Hadamard test and measuring the control qubit in the Pauli-$z$ basis $M$ times allows for an estimation of the cost function value, although it may result in a larger variance compared to direct measurement methods of the $n$-qubit quantum register \cite{Polla2023}. This variance stems from the Hadamard test outcomes being either $+1$ or $-1$, contrasting with direct measurements that yield probability densities across $2^{n}$ distinct basis states for more precise cost function estimations. In Sec. \ref{Sec:Noiseless_Simulations}, we briefly compare the Hadamard test with direct measurement methods, deferring a detailed analysis to future research.

\par From Fig. \ref{Fig:QNPU}(a-c), it is observed that the size of the primary quantum register, $n$, dictating the spatial grid points, $N=2^n$, is distinct from the size of the quantum circuit, which also incorporates additional quantum registers and ancilla qubits for evaluating the energy expectation values. For the NLSE problem, the requisite qubit counts for potential, interaction, and kinetic energy calculations are $2n+1$, $3n+1$, and $2n-1$, respectively, necessitating a minimum of $3n+1$ qubits for the cost function evaluation. Throughout this article, we specify the size of the primary quantum register, $n$, and the minimum qubits needed for the evaluation of the cost function, $n'$, to elucidate both the scale of the problem and the qubit resources essential for the execution of the variational algorithm.

\section{Benchmarking Simulations and Quantum Hardware Implementations}
\label{Sec:Results}

\par In this section, we discuss the results of the variational algorithm, where we solve for the ground state of the time-independent nonlinear Schr\"{o}dinger equation described in Sec. \ref{Sec:NLSE_example}. First, we consider the noiseless (noisy) settings in Sec. \ref{Sec:Noiseless_Simulations} (\ref{Sec:Noisy_Simulations}) and simulate small-size instances of the problem using the Qiskit platform \cite{Anis2021}. In these simulations, we utilize qasm-simulator to compute the cost function values and to determine the trial state probabilities. Since quantum state amplitudes are not directly accessible on a quantum device, the state infidelity is estimated by the residual sum of the state probabilities throughout this work. Then, we turn our attention to digital, gate-based quantum hardware implementation of pre-trained quantum circuits on superconducting IBM Q devices in Sec. \ref{Sec:Device_Demonstration}, where we utilize the ibmq-kolkata and ibmq-mumbai devices. Across all simulations, we consider $0.1$ million shots per circuit and adopt the COBYLA optimizer \cite{Powell1994, Powell1998, Powell2007}. To prepare the trial state, we consider a real-amplitude ansatz shown in Fig. \ref{Fig:HEA_Noiseless}(a). The ansatz consists of $l$ number of layers where each layer consists of a single-qubit $R_{y}$ gate applied to every qubit followed by a sequence of controlled-NOT (CNOT) gates as depicted in Fig. \ref{Fig:HEA_Noiseless}(a). Following the last layer, we apply $R_{y}$ gates to each qubit again such that the ansatz consists of $n(l+1)$ single-qubit gates and variational parameters and $(n-1)l$ CNOT gates. This ansatz transforms the state amplitudes within the $2^{n}$-dimensional real space of the nonlinear problem. 

\par To benchmark the performance of the variational algorithm, we solve the exact ground state $\vert\Psi_{\rm GS}\rangle$ of the NLSE using the imaginary time evolution method \cite{Edwards1995, Dalfovo1999}. Energy of the ground state $\langle\langle{E}\rangle\rangle_{\rm GS}$ is then computed using the ground state amplitudes. This enables us to not only gauge the minimum value of the energy cost function but also to calculate infidelity $F' = \sum_{k=0}^{2^n - 1}\big[ \vert\psi_{{\rm GS}, k}\vert^{2} - \vert\psi_{{\rm var}, k}\vert^2 \big]$, which quantifies the degree to which the trial state probabilities $\vert\psi_{{\rm var},k}\vert^{2}$ resemble those of the ground state.

\subsection{Simulations in Ideal Settings}
\label{Sec:Noiseless_Simulations}

\par  As our method is variational and involves probing the total quantum system indirectly, we first test this approach in simulators. This allows us to distinguish the possible deviation from the ideal results originating from the method itself to the one coming from the quantum hardware noise. To analyze the error arising from the indirect measurement of the quantum system, i.e., the Hadamard test measurement error, we also evaluate the energy expectation value $\langle\langle{E}\rangle\rangle_{\rm Direct}$ - assumed to be exact \footnote{Direct measurement does not involve the QNPU and ancilla qubits. Direct measurement of quantum register has small statistical errors due to shot noise. This error decreases as we increase the number of shots. In our case, we consider $\langle\langle{E}\rangle\rangle_{\rm Direct}$ to be exact, given the large number of shots - $0.1$ million. Therefore, we do not analyze the statistical error in the energy values obtained from the direct measurement method\label{Ref:Footnote1}} - using the direct measurement of the $n$-qubit primary quantum register (see Appendix \ref{App_Sec:Direct_Method} for details). Although $\langle\langle{E}\rangle\rangle_{\rm Direct}$ is measured, the classical optimizer exclusively utilizes the energy expectation value $\langle\langle{E}\rangle\rangle_{\rm var}$ obtained from the Hadamard test measurement, to update the parameters of the quantum ansatz. We define the error as $\Delta_{r, j} = \langle\langle{E}\rangle\rangle^{r, j}_{\rm var} - \langle\langle{E}\rangle\rangle^{r, j}_{\rm Direct}$ for the $j^{th}$ optimizer iteration during the $r^{th}$ execution of the variational algorithm. As $\Delta_{r, j}$ captures the error in the measurement method and is independent of variational parameters, we define an average standard deviation $\sigma = \frac{1}{\sqrt{R{J}}}\sum_{r, j = 1}^{R, J} \sqrt{  \langle \Delta_{r, j} \rangle_{R, J} - \Delta_{r, j}}$, where $\langle\dots\rangle_{R, J}$ denotes the average over the $J$ iterations of the classical optimizer and $R$ executions/realizations of the variational algorithm. $\sigma$ highlights the average spread of the Hadamard test measurement error in the cost function values $\langle\langle{E}\rangle\rangle_{\rm var}$.

\begin{figure}[t]\begin{center}
\includegraphics[clip, trim=0.20cm 4.00cm 0.220cm 0.250cm, width=0.98\linewidth, height=1.40\linewidth, angle=0]{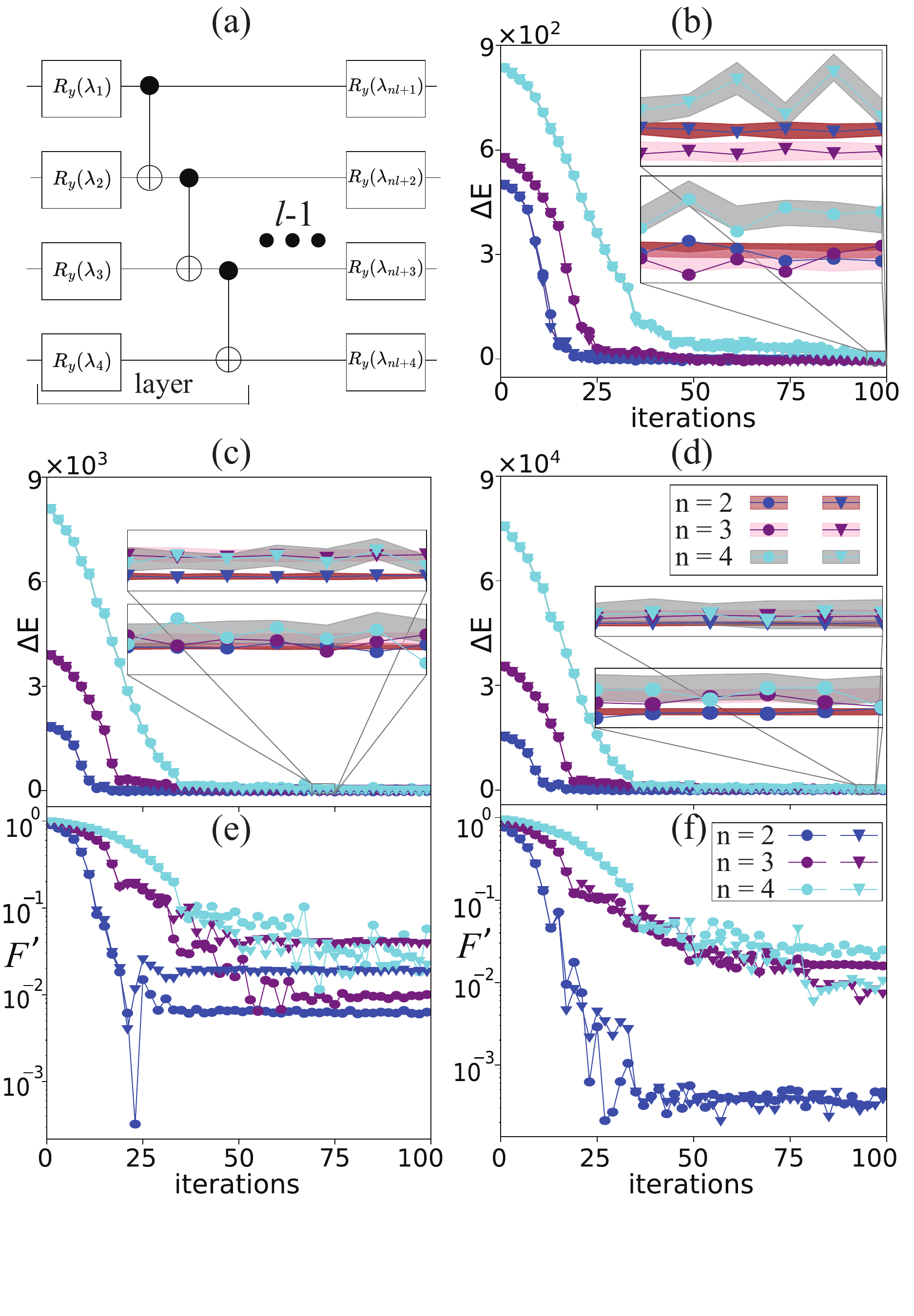}
\caption{Quantum ansatz (a) and results of noiseless simulations: (b-d) cost function values and (e-f) infidelity between the trial and ground state probabilities. Panel (a) shows the real-amplitude ansatz for the case of $n = 4$, where $R_{y}(\lambda_{i})$ is a parameterized single-qubit rotation gate. Energy difference $\Delta{E} = \langle\langle{E}\rangle\rangle_{\rm var} - \langle\langle{E}\rangle\rangle_{\rm GS}$ vs iterations of classical optimizer has been shown for (b) weak nonlinearity $g = 10$, (c) intermediate value of nonlinearity $g = 500$, and (d) strong nonlinearity strength $g = 5000$. Infidelity $F' = \sum_{k=0}^{2^n - 1}\big[ \vert\psi_{{\rm GS}, k}\vert^{2} - \vert\psi_{{\rm var}, k}\vert^2 \big]$ is shown in panel (e) and (f) for $g = 500$ and $g = 5000$, respectively. Here, blue, purple, and cyan color indicates the qubit numbers $n = 2$ ($n' = 7$), $n = 3$ ($n' = 10$), and $n = 4$ ($n' = 13$) for $l = 2$, $l = 4$, and $l = 7$ layers, respectively. Moreover, circular and triangular markers indicate 0.1 million and 2 million shots per circuit, respectively, for the cost function evaluation. \vspace{-0.2cm}}
\label{Fig:HEA_Noiseless}
\end{center}\end{figure}

\par Figs. \ref{Fig:HEA_Noiseless}(b-f) presents the results of noiseless simulations. We examine systems comprising $n = 2$ ($n' = 7$), $n = 3$ ($n' = 10$), and $n = 4$ ($n' = 13$) qubits, depicted by blue, purple, and cyan colors, respectively. Circular markers within these figures indicate the best result among $R = 20$ realizations of the variational algorithm. Additionally, brown, pink, and grey colored regions around the circular markers in Figs. \ref{Fig:HEA_Noiseless}(b - d) highlight the range of average standard deviation $\sigma$ from the $\langle\langle{E}\rangle\rangle_{\rm Direct}$ values. To avoid clutter, $\langle\langle{E}\rangle\rangle_{\rm Direct}$ values are omitted from Figs. \ref{Fig:HEA_Noiseless}(b - d).

\par A comprehensive analysis of the infidelity of trial state probabilities as a function of the number of layers in the real-amplitude ansatz is outlined in Appendix \ref{App_Sec:Layers}. Below, we focus only on a fixed number of layers to illustrate the performance of the variational algorithm. First, considering a relatively weak nonlinearity strength of $g = 10$, the circular markers in Fig. \ref{Fig:HEA_Noiseless}(b) demonstrate the convergence toward the minimum energy, with the energy difference between the variational energy and the ground state energy approaching zero. Second, for intermediate ($g = 500$) and strong ($g = 5000$) nonlinearity strengths, the circular markers in Fig. \ref{Fig:HEA_Noiseless}(c) and \ref{Fig:HEA_Noiseless}(d) depict the convergence to the minimum energy. The infidelity between the ground and trial state probabilities, indicated by circular markers in Figs. \ref{Fig:HEA_Noiseless}(e) and \ref{Fig:HEA_Noiseless}(f) for intermediate and strong nonlinearity strengths, respectively, highlights fidelity exceeding $98\%$ upon convergence. This fidelity may substantially improve by initiating the variational algorithm with an educated guess that possesses a considerable overlap with the ground state. These results highlight that the real-amplitude ansatz efficiently approximates the ground state of the NLSE across different regimes characterized by varying strengths of nonlinearity, thereby demonstrating its expressivity for solving the NLSE with high fidelity.

\par The insets and circular markers in Figs. \ref{Fig:HEA_Noiseless}(b-d) highlight that the range of standard deviation (uncertainty in $\langle\langle{E}\rangle\rangle_{\rm var}$ due to the Hadamard test measurement) increases with increasing system size. Notably, the energy values (refer to circular markers in insets of Figs. \ref{Fig:HEA_Noiseless}(b-d)) exhibit fluctuations that may arise for the following reasons: At a given instance of the classical optimization process, the energy cost function value  $\langle\langle{E}\rangle\rangle_{\rm var}$ incurs an error due to the Hadamard test measurement, is fed to the classical optimizer and results in an erroneous update of the variational parameters. These updated variational parameters, in turn, lead to an energy value with inherent error, resulting in fluctuating behavior near the minimum value. 

\par To reduce the Hadamard test measurement error and provide a more accurate energy cost function value to the classical optimizer, we consider the average value of $R = 20$ executions of each quantum circuit, such that the total number of shots for a given circuit reaches 2 million. In Figs. \ref{Fig:HEA_Noiseless}(b - d), the triangular markers illustrate the difference between the cost function value and the ground state energy $\overline{\langle\langle{E}\rangle\rangle_{\rm var}^{j, r}}^{R} - \langle\langle{E}\rangle\rangle_{\rm GS}$, where $\overline{\cdots}^{R}$ represent average over $R$ executions. Furthermore, we define $\Delta'_{j, r} = \overline{\langle\langle{E}\rangle\rangle_{\rm var}^{j, r}}^{R} - \langle\langle{E}\rangle\rangle_{\rm var}^{j, r}$. The colored region then highlight the span of one standard deviation, $\sigma'_{j} = \frac{1}{\sqrt{R}}\sum_{r = 1}^{R} \sqrt{\langle \Delta'_{j, r}\rangle_{R}  - \Delta'_{j, r} }$, for each iteration $j$ of the classical optimizer. Here $\sigma'_{j}$ describes the spread in the cost function values evaluated using the Hadamard test measurement.  The triangular markers within the insets in Figs. \ref{Fig:HEA_Noiseless}(b - d) reveal a smoother convergence toward the ground state energy while maintaining the same level of fidelity between the ground and final trial state probabilities, as demonstrated in Figs. \ref{Fig:HEA_Noiseless}(e - f). This analysis suggests, as expected, that an increase in the number of shots allows us to reduce the Hadamard test measurement error.

\subsection{Simulations Incorporating Superconducting Quantum Hardware Noise}
\label{Sec:Noisy_Simulations}

\par Before we run the circuits on quantum hardware, we first implement the algorithm in simulators in the presence of realistic noise (see Appendix \ref{App_Sec:Noise_model} for details). The hardware noise features a mean thermal relaxation time ($T_{1}$) of $100\mu{s}$ and a mean dephasing time ($T_{2}$) of $85\mu{s}$, with standard deviations of $30\mu{s}$ and $50\mu{s}$, respectively. It exhibits mean error rates of $2.625\times10^{-4}$ for single-qubit gates and $9.616\times10^{-3}$ for two-qubit gates. Moreover, the mean probability $P_{01}$ ($P_{10}$)
of measuring state $\vert{0}\rangle$ ($\vert{1}\rangle$) when the qubit is prepared in state $\vert{1}\rangle$ ($\vert{0}\rangle$) is $2.06\times10^{-2}$ ($1.98\times10^{-2}$). Here, we assume trivial qubit reset noise, ensuring that each qubit is perfectly initialized in the $\vert{0}\rangle$ state at the onset of each computation. 

\par
\begin{figure}[t]\begin{center}
\includegraphics[clip, trim=0.00cm 0.0cm 0.0cm 0.250cm, width=1.00\linewidth, height=0.95\linewidth, angle=0]{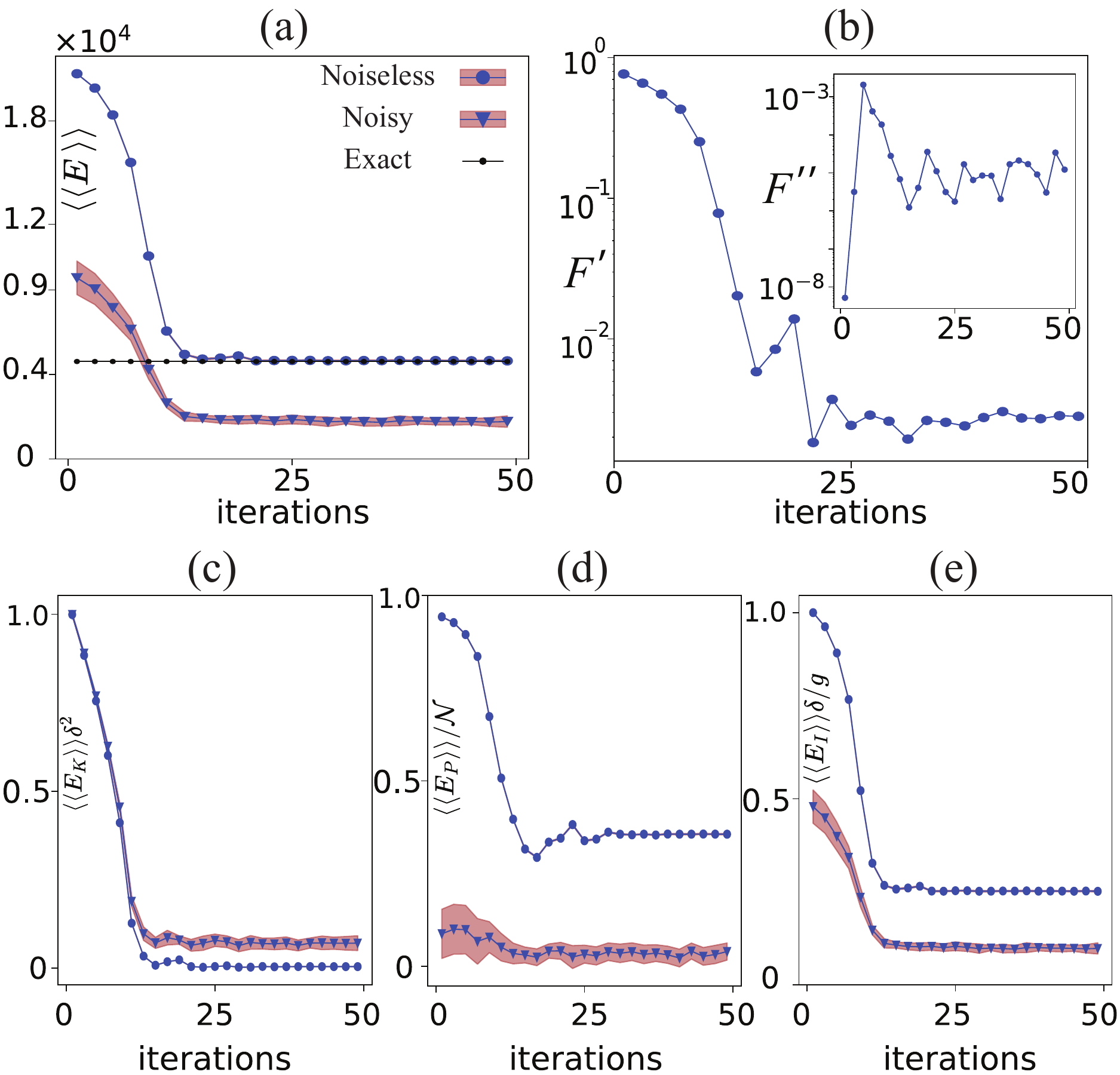}
\caption{Results for noisy simulations: (a) cost function convergence, (b) infidelity of trial state probabilities, and (c-e) effects of hardware noise on cost function components. Circular (triangular) markers in panel (a) shows the energy $\langle\langle{E}\rangle\rangle_{\rm var}$ vs iterations of classical optimizer for $g = 5000$, $n = 2$ ($n' = 7$) and $l = 2$ of real amplitude ansatz in the absence (presence) of quantum noise. Panel (b) (and the inset therein) shows the infidelity $F' = \sum_{k=0}^{2^n - 1}\big[ \vert\psi_{{\rm GS}, k}\vert^{2} - \vert\psi_{{\rm var}, k}^{\rm Noisy}\vert^2 \big]$ ($F'' = \sum_{k=0}^{2^n - 1}\big[ \vert\psi_{{\rm var}, k}^{\rm Noiseless}\vert^{2} - \vert\psi_{{\rm var}, k}^{\rm Noisy}\vert^2 \big]$). Panel (c, d, e) shows the quantities $\langle\langle{E_{K}}\rangle\rangle\delta^{2}$, $\langle\langle{E_{P}}\rangle\rangle/\mathcal{N}$, and $\langle\langle{E_{I}}\rangle\rangle\delta/g$, which are measured from the ancilla qubit. Here, $\delta = 1/2^{n}$ and $\mathcal{N}$ is the norm of the potential function. \vspace{-0.2cm}}
\label{Fig:HEA_Noisy}
\end{center}\end{figure}

 Considering $n = 2~(n' = 7)$ qubit system with strong nonlinearity $g = 5000$, triangular (circular) markers in Fig. \ref{Fig:HEA_Noisy}(a) depict the energy expectation in the presence (absence) of quantum noise. It is important to note that only the energy cost function values obtained in the noisy settings are utilized in the classical optimization to update the variational parameters. Fig. \ref{Fig:HEA_Noisy}(a) highlights that although the energy expectation values from the noisy simulations have smaller magnitudes, they exhibit similar behavior to those obtained in the noiseless settings. Furthermore, Fig. \ref{Fig:HEA_Noisy}(b) illustrates that the final trial state probabilities in the noisy settings closely matches the ground state, indicating that the noisy simulation converges to the set of variational parameters that approximate the ground state of the problem. The measure of infidelity $F'' = \sum_{k=0}^{2^n - 1}\big[ \vert\psi_{{\rm var}, k}^{\rm Noiseless}\vert^{2} - \vert\psi_{{\rm var}, k}^{\rm Noisy}\vert^2 \big]$ between the trial state probabilities generated in noisy and noiseless settings, as depicted in the inset of Fig. \ref{Fig:HEA_Noisy}(b), further suggests that quantum noise scarcely affects the execution of ansatz within the primary quantum register for systems of smaller sizes. 

\par Given the preparation of high-fidelity trial state on the primary quantum register, it is pertinent to consider that quantum noise might influence other distinct processes, such as encoding of the potential function, replication of the variational state, and computation of the energy cost function. To analyze the effect of quantum noise, we examine each component individually, noting the difference in outcomes in the presence and absence of quantum noise. First, we assess the kinetic energy component $\langle\langle{E_{K}}\rangle\rangle/\delta^{2}$, where the corresponding quantum circuit comprises $16$ ($42$) CNOT (single-qubit) quantum gates. As depicted in Fig. \ref{Fig:HEA_Noisy}(c), a minor deviation of approximately $0.05$ is observed in kinetic energy values. With the trial state already prepared to high fidelity, the observed discrepancy in kinetic energy values could be attributed to the impact of quantum gate noise during the computation process.

\par Second, we analyze the potential and interaction energy components, $\langle\langle{E_{P}}\rangle\rangle/\mathcal{N}$ and $\langle\langle{E_{I}}\rangle\rangle\delta/g$, using the corresponding quantum circuits, which incorporate $62$ ($170$) and $133$ ($273$) CNOT (single-qubit) gates, respectively. Notable differences, approximately $0.85$ to $0.35$ and $0.5$ to $0.15$ at the initial and final stages of classical optimization, are observed in Figs. \ref{Fig:HEA_Noisy}(d) and \ref{Fig:HEA_Noisy}(e), reflecting the impact of quantum noise. These disparities in potential and interaction energies likely arise from quantum noise affecting the process of encoding the potential function and replicating the trial states on distinct quantum registers, as well as from the bit-wise multiplication across these registers. For instance, noise impacting the encoding [replication] of the potential function $\hat{V}$ [variational state $\vert\Psi_{\rm var}\rangle$] could result in an entirely different potential function $\hat{V} \pm \hat{\delta{V}}$ [variational state $\vert\Psi'_{\rm var}\rangle$]. 
\begin{figure}[t]\begin{center}
\includegraphics[clip, trim=0.00cm 0.10cm 0.10cm 0.10cm, width=0.95\linewidth, height=0.600\linewidth, angle=0]{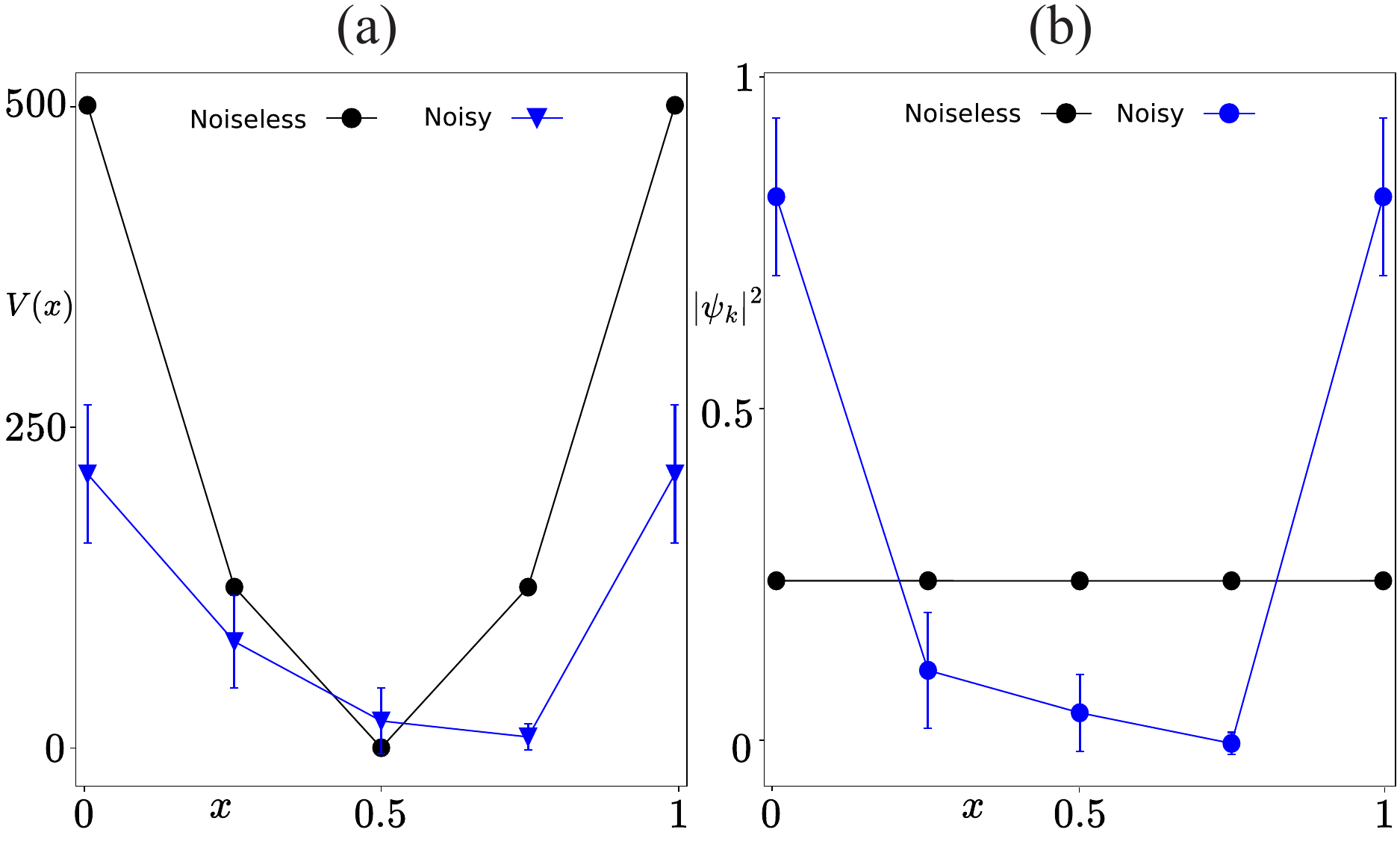}
\caption{ Effect of hardware noise on the encoding of the potential unitary (a) and replication of the variational state (b). Black (blue) indicates the absence (presence) of hardware noise. Error bars represent one standard deviation, calculated from $100$ executions of the noisy circuits. \vspace{-0.2cm}}
\label{Fig:AppendixD}
\end{center}\end{figure}

To verify this, we utilize a system of $1 + n$ qubits, consisting of an ancilla qubit and an $n$-qubit register, which encodes the potential values and/or quantum state. We implement the controlled-$\hat{V}$ and controlled-$U(\lambda)$ operators on the $1 + n$ qubit system, followed by a direct measurement of the $n$-qubit register to extract the potential values and state probabilities. Here, we consider $n = 2$ and $U(\lambda)$ to be the real-amplitude ansatz shown in Fig. \ref{Fig:HEA_Noiseless}(a), with all variational parameters fixed at $\pi/2$. Fig. \ref{Fig:AppendixD} illustrates that hardware noise significantly affects the encoding of potential values and replication of the quantum state, which in turn leads to potential, interaction and total energy values falling below the minimum threshold for the given system, as observed in Figs. \ref{Fig:HEA_Noisy}(a, d-e). Our investigation reveals that deeper quantum circuits with a substantial number of imperfect CNOT and single-qubit gates result in significant deviation and variance in the cost function values, emphasizing the necessity for advanced noise mitigation and/or error correction strategies to improve the quantum computational accuracy and reliability of the variational algorithm.

\subsection{Implementation on Superconducting Cloud Hardware}
\label{Sec:Device_Demonstration}

\par
\begin{figure}[t]\begin{center}
\includegraphics[clip, trim=0.00cm 0.0cm 0.0cm 0.0cm, width=0.90\linewidth, height=1.5\linewidth, angle=0]{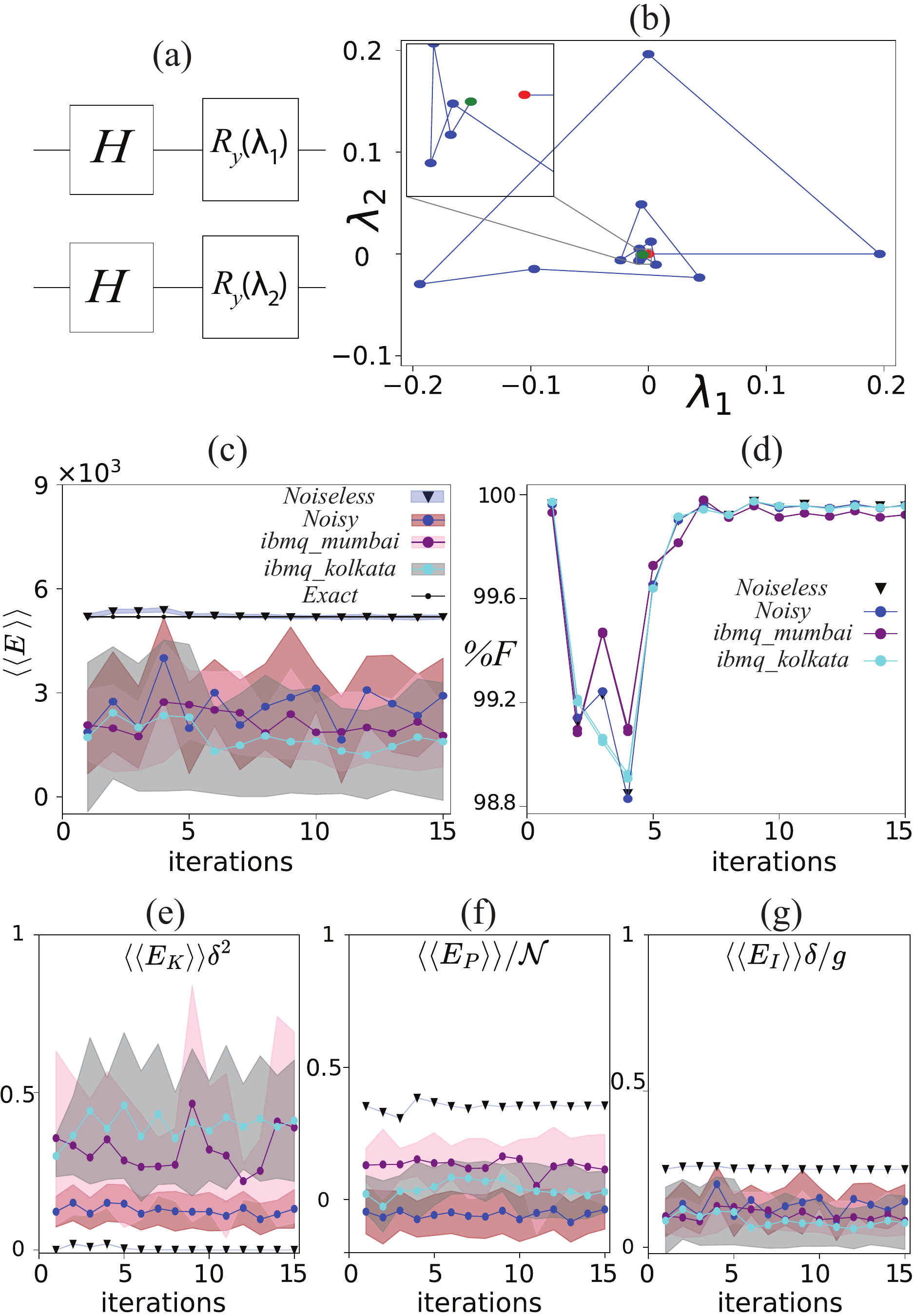}
\caption{The performance of evaluating the cost function on IBM Q devices. Panel (a) shows the simplified quantum ansatz adopted for the quantum device demonstration of the variational algorithm. Panel (b) represent the training of variational parameters, where red (green) marker highlight the initial (final) point of the classical optimization. Panel (c) shows the energy $\langle\langle{E}\rangle\rangle_{\rm var}$ vs iterations of classical optimizer for $g = 5000$, $n = 2$ ($n' = 7$). Here, blue (black) color indicate the noisy (noiseless) simulations, while purple and cyan colors represent simulations performed on ibmq-mumbai and ibmq-kolkata devices, respectively. Moreover, the blue, brown, pink, and gray colors highlight the region of one standard deviation. Panel (d) shows the fidelity $\%F = (1 - \sum_{k=0}^{2^{n}-1}\big[\vert\psi_{\rm GS}\vert^{2} - \vert\psi_{\rm var, k}\vert^{2}\big])\times{100}$ of the trial state probabilities with respect to the ground state solution. Panel (e, f, g) shows the quantities $\langle\langle{E_{K}}\rangle\rangle\delta^{2}$, $\langle\langle{E_{P}}\rangle\rangle/\mathcal{N}$, and $\langle\langle{E_{I}}\rangle\rangle\delta/g$, and corresponding standard deviations. \vspace{-0.2cm}}
\label{Fig:HEA_Device}
\end{center}\end{figure}

 We now implement pre-trained circuits on the digital gate-based quantum devices, ibmq-kolkata and ibmq-mumbai \cite{IBMQ}, both of which feature identical topology and basis gate sets. These two devices were selected due to their availability and our limited access to IBMQ processors. We focus on $n = 2$ ($n' = 7$) qubit system with $g = 5000$ and design the quantum ansatz tailored to the strong nonlinearity case, such that each qubit has a Hadamard gate followed by a layer of parameterized single-qubit $R_{y}$ rotation gate, as shown in Fig. \ref{Fig:HEA_Device}(a). It is important to highlight that our goal here is to analyze the performance of evaluating the cost function on the quantum devices; therefore, we restrict the variational space, which might not include the exact ground state. The simplified ansatz of Fig. \ref{Fig:HEA_Device}(a) offers two advantages. The first advantage is the absence of CNOT gates, resulting in quantum circuits with fewer entangling gates and a shallow circuit depth. Consequently, the quantum circuits to measure kinetic, potential, and interaction energies consist of $14$, $59$, and $70$ ($33$, $160$, and $124$) CNOT (single-qubit) gates, respectively. The second advantage is that, for the zero value of each variational parameter, the quantum ansatz generates a trial state with over $99\%$ fidelity with the exact ground state in the strong nonlinearity regime. This insight allows us to restrict the variational space closer to the ground state, such that even for the non-zero but smaller values of the variational parameters, the trial state maintains considerable overlap with the ground state. With this setting, we execute the variational algorithm in the presence of hardware noise, where each parameter of the ansatz is initiated at zero value (red point in Fig. \ref{Fig:HEA_Device}(b)). The classical optimizer explores the two-dimensional variational space for a few iterations before converging toward the zero values of the variational parameters (a green point in Fig. \ref{Fig:HEA_Device}(b) indicates the set of final values of the variational parameters).

\par With these pre-trained quantum circuits, we measure the energy cost function and fidelity of state probabilities in both noiseless simulations and noisy settings of simulations and digital quantum hardware. It is worth noting that, unlike the noisy simulations, quantum hardware simulations exhibit qubit reset noise. Fig. \ref{Fig:HEA_Device}(d) demonstrates the percentage fidelity $\%F = (1 - \sum_{k=0}^{2^{n}-1}\big[\vert\psi_{\rm GS}\vert^{2} - \vert\psi_{\rm var, k}\vert^{2}\big])\times{100}$ of the trial state probabilities with a maximum disparity of $0.25\%$ between noiseless simulations and quantum device simulations, highlighting the high-fidelity preparation of the trial state across the two devices. Additionally, the evaluation of the energy cost function and components in both noiseless (depicted in black) and noisy (depicted in blue) simulations, as shown in Figs. \ref{Fig:HEA_Device}(c), \ref{Fig:HEA_Device}(e-g), aligns with the findings presented in Sec. \ref{Sec:Noisy_Simulations}, with differences and variances stemming from the impact of quantum noise.

\par Figs. \ref{Fig:HEA_Device}(c) and \ref{Fig:HEA_Device}(e-g) show the energy cost function and individual components measured on the ibmq-mumbai (in purple color) and ibmq-kolkata (in cyan color) devices. The results exhibit behavior akin to those observed in the noisy simulations. Here, the qubit reset noise, causing imperfect initialization of ancilla qubits and quantum registers, further impacts the encoding (preparation) of the potential function (variational state) and the execution of the adder circuit, resulting in significant increases in standard deviation values. These findings reveal large errors and variances, thereby highlighting the limitations of the current NISQ devices in executing the variational algorithm for nonlinear cost functions.

\section{Summary and Outlook}\label{Sec_Conclusion}

\par In this work, we studied the ground state problem of the nonlinear Schr\"{o}dinger equation by utilizing the variational quantum algorithm. For a quadratic potential, we demonstrated that the real-amplitude ansatz, which spans the $2^n$-dimensional real space of the problem, has the expressivity to represent the ground state in the weak, intermediate, and strong regimes of nonlinearity. Furthermore, we analyzed the Hadamard test measurement error against a direct method and observed that the Hadamard test measurement error results in fluctuations in the cost function, potentially leading to optimization challenges. These fluctuations in the cost function can be averaged out by repeated measurements of the same quantum circuit, thus providing a more stable and reliable basis for the optimization process. 

\par Secondly, we incorporated quantum hardware noise into the simulations and reported that while the variational algorithm produces a high-fidelity ground state for the small system, the values of the cost function and its constituting components are significantly influenced by hardware noise. We argued that, given the high-fidelity trial state, quantum noise affects the cost function computation during the execution of the quantum nonlinear processing unit (QNPU). Finally, we implemented pre-trained circuits on the ibmq-kolkata and ibmq-mumbai devices and observed that qubit reset noise, which was not considered in the noisy simulations, further affects the evaluation of the cost function. Our results highlight the limitations of the current NISQ devices for the implementation of the variational algorithm for nonlinear problems. 

\par In the future, an extensive analysis of the Hadamard test measurement error might be an interesting aspect to pursue, especially where it is critical to explore how the Hadamard test measurement error scales with the number of qubits, depth of the quantum circuit, and the number of entangling gates. Furthermore, studying noise mitigation and error correction techniques to improve the cost function evaluation in the variational algorithm would be intriguing. Lastly, the investigation of fluid dynamics problems, modeled by the one-dimensional Burgers' equation and the two- and three-dimensional Navier-Stokes equations, is worthwhile, where a few key aspects to analyze may include the expressivity of various ansatzes, challenges of classical optimization, and the impact of quantum noise, to name a few. 

\acknowledgments
MU and EM contributed equally. We acknowledge Nis Van H\"{u}lst and Pia Siegl for helpful discussions on the theory and computation of tensor networks. This research is supported by the EU HORIZON - Project 101080085 - QCFD, the National Research Foundation, Singapore and A*STAR under its CQT Bridging Grant, and Quantum Engineering Programme NRF2021-QEP2-02-P02. We thank IBM Quantum for the cloud quantum computing access.

\appendix
\section{Construction of Potential Unitary $\hat{V}$}
\label{App_Sec:MPS}

\par In this section, we review the procedure of encoding the potential function into a unitary matrix $\hat{V}$ such that $V_{k} = \mathcal{N}\langle{\rm binary}(k)\vert\hat{V}\vert{0}\rangle = \mathcal{N}\langle{\rm binary}(k)\vert{\psi_{P}}\rangle$, where $\mathcal{N}$ is the norm of the potential function, $\vert{\rm binary}(k)\rangle$ is the $k^{th}$ basis state, and
\bea\bal\nonumber
\vert\psi_{P}\rangle = \hat{V}\vert{0}\rangle = \sum_{i_1,\dots,i_n}V_{i_1,\dots,i_n}\vert i_1,\dots,i_n \rangle
\eal\eea
is the state that we intend to prepare on a separate quantum register (refer to Fig. \ref{Fig:QNPU}(a)). Here, $\vert i_1, \dots, i_n\rangle = \vert{\rm binary}(k)\rangle$ are the $n$-qubit basis states, $V_{i_1,\dots,i_n} = V_{k}/\mathcal{N}$ are the potential values or probability amplitudes of the state, and $i'$s are the physical indices representing qubits. For this purpose, we utilize the framework of matrix product states (MPS) \cite{Verstraete2008, Oseledets2013, Orus2014, Cirac2021} and transform the state $\vert{\psi_{P}}\rangle$ in the form 
\bea\bal
\vert\psi_{P}\rangle = \sum_{i_1,\dots,i_n}A_{i_1,r_1}A_{r_1,i_2,r_2}\dots&~{A_{r_{n-2},i_{n-1},r_{n-1}}}\\&~~~~~~~~A_{r_{n-1},i_n}\vert i_1,\dots,i_n \rangle,
\label{EQ:MPS_decomposition}
\eal\eea
where the first and last tensors are of rank-$2$, the middle tensors are of rank-$3$, and $r'$s represent dummy indices that control the bond-dimension of the matrix product states \cite{Orus2014, Cirac2021}. Here, we have considered the Einstein summation convention where repeated indices are summed over. Moreover, without loss of generality, we consider each ($i$) $r$ to be (two-) $2^{j}$-dimensional for $j \in \{1, 2, \dots, n-1\}$.

\par Given a high-rank tensor $V_{i_1,\dots,i_n}$, we perform successive operations: reshaping, where a high-rank tensor is reshaped into a rank-$2$ tensor; singular value decomposition (SVD) $M = \mathcal{S}\mathcal{V}\mathcal{D}^{T}$, where a rank-$2$ tensor is decomposed into matrices of singular values and corresponding right and left eigenvectors; and tensor contraction, where two or more tensors are contracted to form a single tensor. Here, singular values of matrix $M$ are arranged in descending order in the diagonal of the $\mathcal{V}$-matrix, and left (right) eigenvectors of $M$ form the columns of matrix $\mathcal{S}$ ($\mathcal{D}$). The general algorithm to obtain the MPS format of Eq. (\ref{EQ:MPS_decomposition}) is as follows. 

\begin{algorithm}[H]
\caption{MPS Algorithm}\label{MPS-Algorithm}
{\bf Input:} High rank tensor\;
\While{Number of tensors $<$ Number of qubits}{
    Reshape tensor into matrix: $V_{i_1, \dots, i_n} \rightarrow M_{i_1, i_{2}\dots{i_n}}$ \;
    Perform SVD decomposition of matrix: $M = \mathcal{S}\mathcal{V}\mathcal{D}^{T}$ \;
    Contract $\mathcal{V}$ and $\mathcal{D}^{T}$: $M^{'} = \mathcal{V}\mathcal{D}^{T}$ \;
    Update: $V = M^{'}$ \;
    }
Reshape isometries to appropriate rank \;
{\bf return} MPS Format. 
\end{algorithm}

\par
\begin{figure}[t]\begin{center}
\includegraphics[clip, trim=0.0cm 0.00cm 0.0cm 0.0cm, width=1.00\linewidth, height=0.70\linewidth, angle=0]{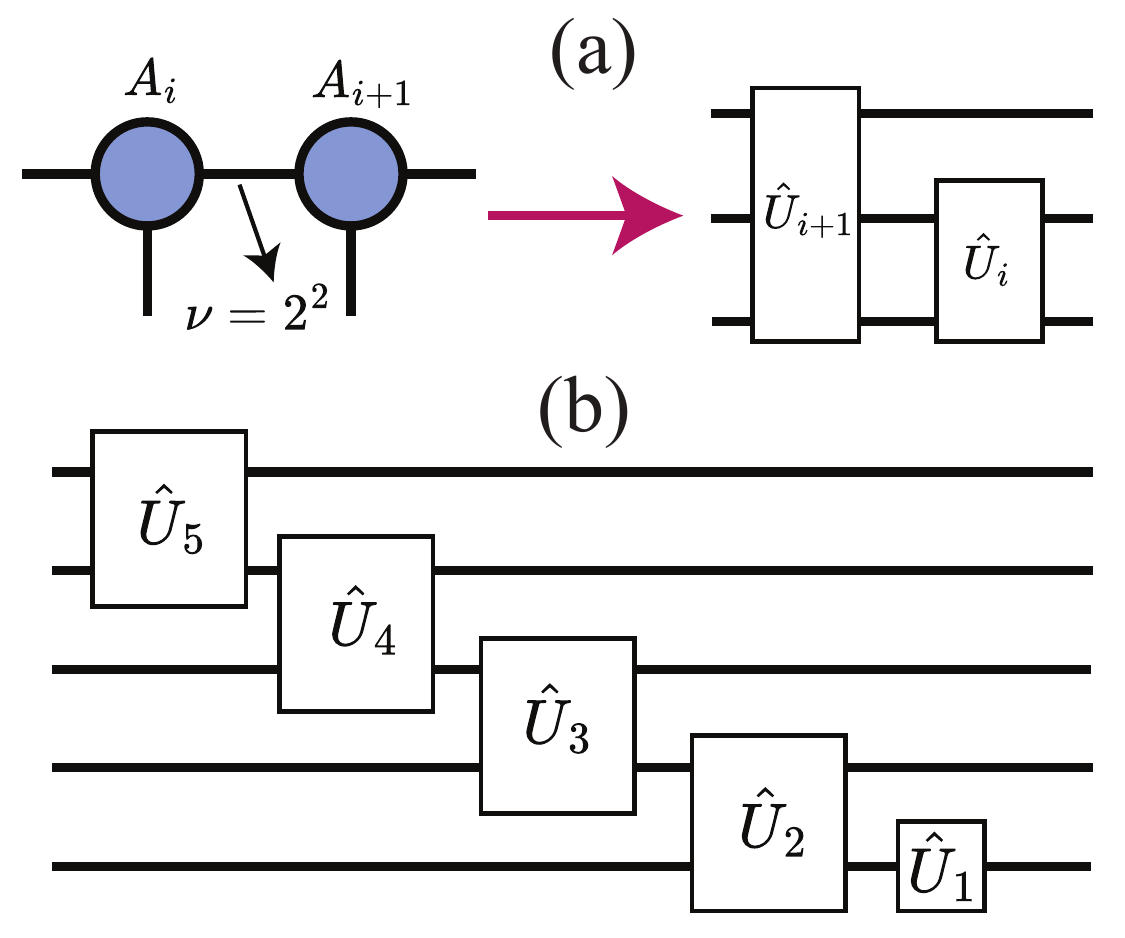}
\caption{ Depiction of tensor network to quantum circuit mapping. Panel (a) illustrates the transformation of a small tensor network into a unitary circuit, highlighting that the bond dimension of $2^{j}$ between two adjacent tensors results in $j$ overlapping qubits for the two unitary gates. Schematic of a quantum circuit obtained from a tensor network with a maximum bond dimension of $2$ is shown in panel (b). \vspace{-0.2cm}}
\label{Fig:MPS}
\end{center}\end{figure}

 With a maximum value of bond dimension $\nu = 2^{\kappa}$, the MPS in the left canonical form is written as 
\bea\bal\nonumber
&V_{i_1,\dots,i_n}\\ &~~~~= \sum_{\substack{r_j=1 \\ j=1,\dots,n-1}}^D {A}_{i_1,r_1}{A}_{r_1,i_2,r_2}\dots&{A}_{r_{n-2},i_{n-1},r_{n-1}}{A}_{r_{n-1},i_n}
\eal\eea
where, 
\bea\nonumber
D = \left\{
\begin{array}{ll}
      2^j,&~~~ if ~~~2^j \le \nu \\
      \nu,&~~~ if ~~~ 2^j > \nu.
\end{array} 
\right. 
\eea
Here, $\nu = 2^{\kappa}$ determines the accuracy of the approximation of the matrix product states (MPS) and bond dimensions between adjacent tensors are $2, 2^2,..., 2^{\kappa-1}, 2^\kappa,..., 2^\kappa, 2^{\kappa-1},..., 2^2, 2$. The contraction of two adjacent tensors with a common bond dimension of $2^{j}$ is represented by $j$ overlapping qubits between the two unitary gates, as shown in Fig. \ref{Fig:MPS}(a).

\par To transform the MPS format into unitary gates, we follow the procedure described in Ref. \cite{Lubasch2020}. For the given MPS format, we first contract all the leftmost tensors up to the bond dimension $2^{\kappa-1}$. The MPS representation then comprises of $n-(\log_2(v)-1)$ tensors with bond dimensions ranging from $2^\kappa$ to $2$, between adjacent tensors. Given the compact MPS representation, each tensor contains fewer elements, which are insufficient for creating the appropriate gates. To address this issue, we extend each tensor by combining it with its nullspace and adding an additional index, thus compensating for the missing elements needed for the suitable gate. Each middle tensor is transformed into a rank-$4$, while the leftmost and rightmost tensors are transformed into rank-$3$ tensors. Finally, we reshape each extended tensor into a matrix, carefully maintaining the qubit ordering. The placement of each element within the unitary is critical, as it corresponds to different qubits in the quantum circuit. A generic algorithm for transforming the MPS format into a quantum circuit is as follows.
\begin{algorithm}[H]
\caption{MPS to Quantum Circuit}\label{MPStoQuantumCircuit}
\textbf{Input:} MPS format\;
Contract tensors up to bond dimension \(2^{\kappa-1}\)\;
\For{each remaining tensor}{
    Reshape rank-$3$ tensor into matrix: $A_{r_{j-1},i_j,r_j} \xrightarrow{} M_{r_{j-1}i_j,r_j}$~\;
    Stack the tensor with its nullspace by introducing an extra index~\;
    Reshape the extended tensor into a matrix~\;
    Rearrange indices according to qubit ordering convention~\;
    }
Apply resulting unitaries to a quantum circuit~\;
\Return{Quantum Circuit}\;
\end{algorithm}
\noindent An example of a quantum circuit representing a generic MPS of bond dimension $2$ is shown in Fig. \ref{Fig:MPS}(b).

\section{Direct Measurement of Cost Function}
\label{App_Sec:Direct_Method}
In this section, we describe a direct measurement method of evaluating the cost function discussed in Ref. \cite{Lubasch2020}, which does not involve a quantum nonlinear processing unit (QNPU), ancilla qubits, and the Hadamard test measurement. We use this direct measurement method to validate the cost function evaluation of the variational algorithm.

\begin{figure}[hbt]
\begin{center}
\includegraphics[clip, trim=0.30cm 0.40cm 0.400cm 0.10cm, width=0.90\linewidth, height=0.65\linewidth, angle=0]{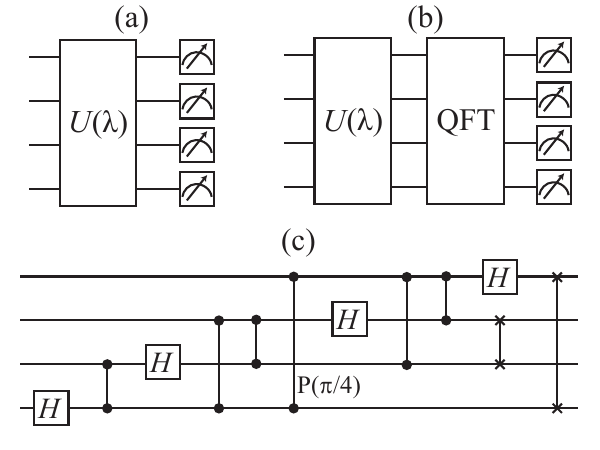}
\caption{Direct measurement method: (a-b) quantum circuits to obtain the energy components, and (c) the quantum Fourier transform circuit. Panel (a) depicts the quantum circuit to calculate the probabilities of the final state. These probabilities is used to calculate the expectation value of potential and interaction energies. Panel (b) depicts the quantum circuit used to construct the quantum state $\vert\Psi'\rangle = {\rm QFT}\vert\Psi\rangle$ in the basis of the Laplace operator. Measuring all the qubits give probabilities $\vert\psi'_{k}\vert^{2}$. Panel (c) shows the decomposition of the quantum Fourier transform operation, where $H$ is the Hadamard gate and $P(\pi/4)$ is the controlled-phase gate. \vspace{-0.2cm}}
\label{Fig:Direct_Method}
\end{center}
\end{figure}

We consider an $n$-qubit quantum register such that it describes the nonlinear problem on $N = 2^{n}$ grid points. This quantum register is initialized in the fixed $\vert{0}\rangle^{\otimes{n}}$ state. The initial state is then transformed into some final state by applying the quantum ansatz $U(\lambda)$ considered in the variational algorithm. As the final step, we measure all the qubits, as shown in Fig. \ref{Fig:Direct_Method}(a), and obtain the probability outcome $\vert\psi_{k}\vert^{2}$ associated with each basis state $\vert{\rm binary}(k)\rangle$. The probability density is then plugged into Eq. (\ref{EQ:Cost_Functions}) to compute the expectation values of potential energy $\langle\langle{E_{P}}\rangle\rangle = \sum_{k}\vert\psi_{k}\vert^{2}V_{k}$ and interaction energy $\langle\langle{E_{I}}\rangle\rangle = \sum_{k}\vert\psi_{k}\vert^{m + 2}$ on a classical computer.

To evaluate the expectation value of kinetic energy, we transform the final state using a quantum Fourier transform (QFT) circuit \cite{Lubasch2018}, which also diagonalizes the Laplace operator. This way we represent the final state in the basis of the Laplace operator. The expectation value of kinetic energy is then given as $\langle\langle{E_{K}}\rangle\rangle = \frac{1}{2}\langle\psi\vert{\rm (QFT)}^{\dagger}\hat{D}_{\Delta}{\rm (QFT)}\vert\psi\rangle = \frac{1}{2}\sum_{k}\vert\psi^{'}_{k}\vert^{2}\Delta_{k}$, where $\hat{D}_{\Delta}$ has $\Delta_{k}$ values along the diagonal, and $\Delta_{k} = 2^{2n}\big[\cos(2\pi{k}/2^{n}) - 1\big]$ are the eigenvalues of the Laplace operator. Here, $\vert\psi^{'}_{k}\vert^{2}$ is the probability of the $k^{th}$ basis state obtained by measuring all the qubits after applying quantum Fourier transform to the final state as depicted in Fig. \ref{Fig:Direct_Method}(b-c). We obtain the expectation value of total energy by summing up individual components $\langle\langle{E_{Direct}}\rangle\rangle = \langle\langle{E_{P}}\rangle\rangle + \langle\langle{E_{I}}\rangle\rangle + \langle\langle{E_{K}}\rangle\rangle$. Here, $\langle\langle{E_{Direct}}\rangle\rangle$ is the energy expectation value obtained using the direct method, as discussed above. In the absence of any noise, direct measurement method gives the correct expectation values of kinetic, potential, and interaction energies. 

\section{Number of Layers in Real Amplitude Ansatz}
\label{App_Sec:Layers}

\begin{figure}[t]\begin{center}
\includegraphics[clip, trim=0.00cm 0.10cm 0.10cm 0.10cm, width=0.95\linewidth, height=0.600\linewidth, angle=0]{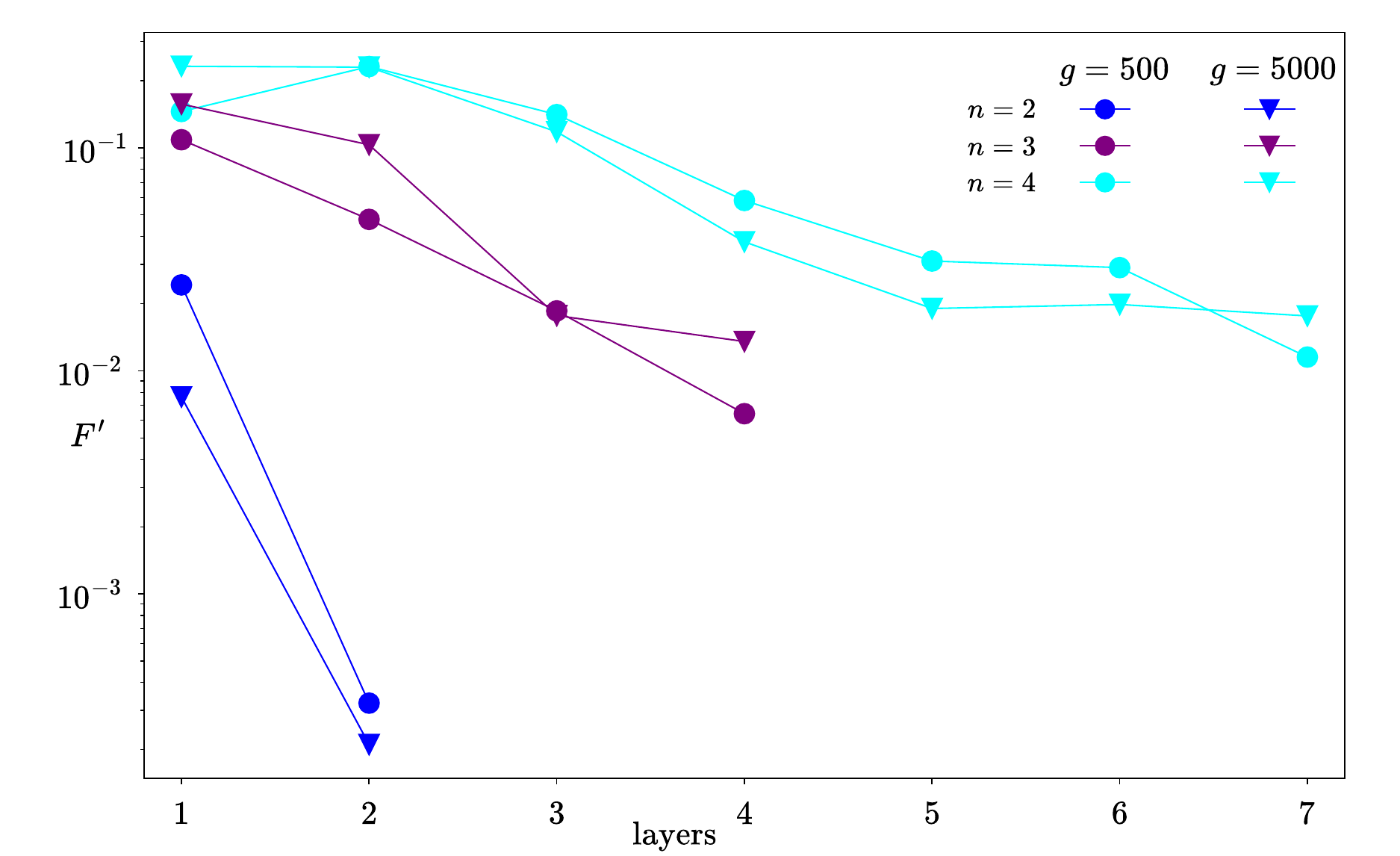}
\caption{Convergence to ground state as a function of number of layers in real-amplitude ansatz. Blue, purple, and cyan colors indicates qubit numbers $n = 2$, $n = 3$, and $n = 4$, and circular (triangular) markers represent the strength of nonlinearity $g = 500$ ($g = 5000$). \vspace{-0.2cm}}
\label{Fig:AppendixC}
\end{center}\end{figure}

In this section, we analyze the convergence of the variational algorithm as a function of the number of layers in the quantum ansatz. To this end, we execute the quantum circuits outlined in Sec. \ref{Sec:NLSE_example} and minimize the energy cost function. Although we minimize the energy of the system, which inherently contains Hadamard test measurement errors, we adopt the infidelity of the trial state probabilities as a measure to analyze convergence. Fig. \ref{Fig:AppendixC} highlights that, as we increase the number of layers in the ansatz, the infidelity between the ground and final trial state probabilities decreases. This suggests that larger problems will require more layers in the real-amplitude ansatz to adequately express the solution to the ground state problem of the NLSE.

\section{Noisy Simulations}
\label{App_Sec:Noise_model}
\par Various types of noise errors have been identified that detrimentally impact the computational performance of quantum hardware. These sources encompass reset errors, wherein the qubit is initialized in an imperfect state; measurement errors, wherein the qubit's state (either $\vert{0}\rangle$ or $\vert{1}\rangle$) is inaccurately measured ($\vert{1}\rangle$ or $\vert{0}\rangle$, respectively); and gate errors, wherein interactions with the environment induce irregularities in the outcomes of quantum operations.

\par To model the effect of quantum hardware noise within our simulations, we prepare a qasm-simulator tailored to the noise properties of the ibmq-kolkata device and execute the simulations on this noisy qasm-simulator. The device properties include relaxation and coherence times ($T_{1}$ and $T_{2}$) for each qubit, gate duration ($\tau$) for each single- and two-qubit gate, and probabilities $P_{01}$ ($P_{10}$) of measuring state $\vert{0}\rangle$ ($\vert{1}\rangle$) when the qubit is prepared in state $\vert{1}\rangle$ ($\vert{0}\rangle$). Given that the device calibration data does not contain any information on the qubit reset error, we neglect the effect of this error in the noisy simulations. Subsequently, we utilize the calibration data within Qiskit's integrated functions to configure a noisy qasm-simulator. Moreover, we incorporate the device's coupling map and transpile each quantum circuit to the basis gate set of the ibmq-kolkata device. This methodology yields an approximate noise model, albeit one that does not encapsulate all potential sources of noise errors inherent to the quantum devices. In the noisy simulations presented in this work, we utilized the ibmq-kolkata device calibration data obtained at $05:37$ GMT on November $28$, $2023$.

\bibliographystyle{apsrev4-2}
\bibliography{References_QCFD}

\end{document}